\documentclass[twocolumn,showpacs,amsmath,amssymb]{revtex4}
\usepackage{graphicx}
\usepackage{dcolumn}
\usepackage{bm}

\newcommand{\ba}{\begin{eqnarray}}
\newcommand{\ea}{\end{eqnarray}}
\newcommand{\ban}{\begin{eqnarray*}}
\newcommand{\ean}{\end{eqnarray*}}
\newcommand{\be}{\begin{equation}}
\newcommand{\ee}{\end{equation}}
\newcommand{\bd}{\begin{displaymath}}
\newcommand{\ed}{\end{displaymath}}

\newcommand{\n}[1]{\label{#1}}

\newcommand{\eq}[1]{(\ref{#1})}

\newcommand{\E}{{\mathbb E}}

\newcommand{\pa}{\partial}

\newcommand{\hh}{\, ,\hspace{0.5cm}}
\newcommand{\hhh}{\, ,\hspace{0.2cm}}

\newcommand{\pr}[5] {\bibitem{#1} {#2}, Phys. Rev.  {\bf D#3}, #4 (#5).}
\newcommand{\cqg}[5] {\bibitem{#1} {#2},  Class. Quant. Grav.  {\bf #3}, #4 (#5).}

\begin{document}

\draft


\title{Embedding of the Kerr-Newman Black Hole Surface in 
Euclidean Space}
\author{Valeri P. Frolov}
\email{frolov@phys.ualberta.ca}
\affiliation{$^{1}$Theoretical Physics Institute, University of Alberta,
  Edmonton, AB, Canada, T6G 2J1}

\date{\today}

\begin{abstract}
We obtain a global embedding of the surface of a rapidly rotating
Kerr-Newman black hole in an Euclidean 4-dimensional space.
\end{abstract}

\pacs{04.70.Bw, 04.50.+h, 04.20.Jb \hfill Alberta-Thy-01-06}

\maketitle

\section{Introduction}

In this paper we discuss the problem of  isometric embedding of
the surface of a rapidly rotating black hole in a flat space.

It is well known that intrinsically defined Riemannian manifolds can
be isometrically embedded in a flat space. According to the
Cartan-Janet \cite{jan,car} theorem, every analytic Riemannian manifold of
dimension $n$ can be locally real analytically isometrically embedded
into $\E^N$ with $N={n(n+1)/2}$.  The so called Fundamental Theorem
of Riemannian geometry  (Nash, 1956 \cite{nash}) states that every
smooth Riemannian manifold of dimension $n$ can be globally
isometrically embedded in a Euclidean space $\E^N$ with
$N={(n+2)(n+3)/2}$.

\begin{figure}[tp]
\begin{center}
\includegraphics[height=2.5cm,width=4cm]{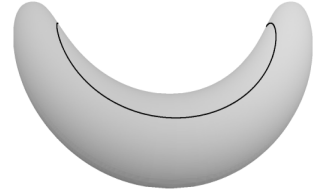}
\caption{This picture shows the "croissant" surface. A solid line
separates two regions with opposite signs of the Gauss curvature.
Each of these regions has the topology of a 2D ball. The Gauss curvature
is negative in the upper ball-region.}
\label{f1}
\end{center}
\end{figure}

The problem of isometric embedding of 2D manifolds in $\E^3$ is well
studied. It is known that any compact surface embedded isometrically
in $\E^3$ has at least one point of positive Gauss curvature. Any 2D
compact surface with positive Gauss curvature is always isometrically
embeddable in $\E^3$, and this embedding is unique up to rigid
rotations. (For general discussion of these results and for further
references, see e.g. \cite{ber}). It is possible to construct
examples when a smooth geometry on a 2D ball with negative Gauss
curvature cannot be isometrically embedded in $\E^3$ (see e.g.
\cite{pog,ny}). On the other hand, it is easy to construct an example
of a global smooth isometric embedding for a surface of the topology
$S^2$ which has both, positive and negative Gauss curvature 
ball-regions, separated by a closed loop where the Gauss curvature
vanishes. An example of such an embedding is shown in Figure~\ref{f1}
\cite{pol}.

The surface geometry of a charged rotating black hole and its
isometric embedding in $\E^3$ was studied long time ago by Smarr
\cite{Smarr}.  He showed that when the dimensionless rotation
parameter $\alpha=J/M^2$ is sufficiently large,
there are two regions near poles of the horizon surface where the
Gauss curvature becomes negative. Smarr proved that these regions
cannot be isometrically embedded (even locally) in $\E^3$ as a
revolution surface, but such local embedding is possible in a 3D
Minkowsky space. More recently different aspects of the embedding of
a surface of a rotating black hole and its ergosphere in $\E^3$ were
discussed in \cite{pnl,bas}. A numerical scheme for construction of
the isometric embedding for surfaces with spherical topology was
proposed in \cite{bas}. The surface geometry of a rotating black hole
in an external magnetic field and its  embedding in $\E^3$ was
studied in \cite{wk,wkd,kd}.

The purpose of this paper is to obtain the global isometric embedding
of a surface of a rapidly rotating black hole in $\E^4$. In Section~2
we discuss general properties of 2D axisymmetric metrics and prove
that if the Gauss curvature is negative at the fixed points of the
rotation group it is impossible to isometrically embed a  region
containing such a fixed point in $\E^3$. In Section~3 we demonstrate
that  such surfaces can be globally embedded in $\E^4$. We obtain the
embedding of  surfaces of rapidly rotating black holes in $\E^4$ in
an explicit form in Section~4. Section~5 contains a brief summary and
discussions.

\section{Geometry of 2D axisymmetric distorted spheres}

Let us consider an axisymmetric deformation $S$ of a unit sphere $S^2$.
Its  metric can be written in the form
\be\n{m1}
dl^2=h(x)dx^2+f(x)d\phi^2\, .
\ee
Here $\xi=\pa_{\phi}$ is a Killing vector field with closed trajectories.
Introducing a new coordinate $\mu=\int dx \sqrt{h f}$ one can rewrite 
\eq{m1} in the form
\be\n{m}
dl^2=f(\mu)^{-1}d\mu^2+f(\mu) \, d\phi^2\, .
\ee
We assume that the function $f$ is positive inside the interval
$(\mu_0,\mu_1)$ and vanishes at its ends. We choose $\mu_0=-1$. The
surface area of $S$ is $2\pi (\mu_1+1)$. By  multiplying the metric
\eq{m1} by a constant scale factor $(\mu_1+1)^{-1/2}$ one can always
put $\mu_1=1$, We shall use this choice, for which the surface area
of $S$ is $4\pi$ and the fixed points of $\xi$ are located at
$\mu=\pm 1$. 

The metric \eq{m} is regular (no conical singularities) at the 
points $\mu=\pm 1$ (where $r=0$) if $f'(\pm 1)=\mp 2$. (Here and
later $(\ldots)'=d(\ldots)/dz$.) The Gaussian curvature for the
metric \eq{m} is
\be
K=-{1\over 2}f''\, .
\ee

Let us introduce a new coordinate $r=\sqrt{f(\mu)}$. The metric
\eq{m} in these coordinates is 
\be \n{VV}
dl^2=V(r)dr^2+r^2 d\phi^2\hh
V={4\over {f'}^2}\, . 
\ee 
We denote by $K_0$ the value of $K$ at a fixed point of rotations.
Then
in the vicinity of this point one has
\be\n{VVV}
V\approx 1+{1\over 2}K_0 r^2 +\ldots\, .
\ee
For $K_0<0$ the region in the vicinity of the fixed point cannot be
embedded as a revolution surface in a Euclidean space $\E^{n}$ for any
$n\ge 3$. Indeed, consider a space
$\E^{n}$ with the metric 
\be
dS^2=dX^2+dY^2+\sum_{i=3}^n dZ_{i}^2\, .
\ee
For the surface of revolution $X=r \cos \phi$, $Y=r\sin\phi$, 
$Z_i=Z_i(r)$ the induced metric is
\be
dl^2=V(r) dr^2+r^2 d\phi^2\hh V(r)=1+\sum_{i=3}^n (dZ_i/dr)^2\, .
\ee
For a regular surface  $V(0)=1$ and $V(r)\ge 1$ in the vicinity of
$r=0$. According to \eq{VVV} this is impossible when $K_0<0$. 

We show now that if $K_0<0$ then a ball-region near the fixed point
$p_0$ of axisymmetric 2D geometry cannot be isometrically embedded in
$\E^3$. Let us assume that such an embedding (not necessarily as a
revolution surface) exists. One can choose coordinates $(X^1,X^2,Z)$ in
$\E^3$ so that $X^1=X^2=0$ at $p_0$, and in its vicinity
\be
Z={1\over 2}(k_1 {X^1}^2+k_2 {X^2}^2)+\ldots\, ,
\ee
where $k_a$ $ (a=1,2)$ are principal curvatures at $p_0$.  Here and
later `dots' denote omitted higher order terms. The metric on this
surface induced by its embedding is
\[
dl^2= (1+k_1^2 {X^1}^2) d{X^1}^2 +
(1+k_2^2 {X^2}^2) d{X^2}^2
\]
\be\n{im}
+ 2k_1 k_2 {X^1}{X^2}d{X^1}d{X^2} +\ldots \, .
\ee
In the vicinity of $p_0$ the Killing vector $\xi$ generating rotations
has the form
\be\n{k}
\xi= p^a\pa_{a}\, ,
\ee
where $p^a$ $ (a=1,2)$ are regular functions of $(X^1,X^2)$ vanishing at
$(0,0)$. Their expansion near $p_0$ has the form
\be\n{S}
p^a=P^a_b X^b + P^a_{bc} X^b X^c + P^a_{bcd} X^b X^c X^d+\ldots .
\ee
Consider the Taylor expansion  near $p_0$ of the Killing equation 
\be\n{kil}
\xi_{a:b}=\xi_{a,b}-\Gamma_{ab}^{\ \ c}\xi_c=0
\ee
in the metric \eq{im}. Since the expansion of both $\Gamma_{ab}^{\ \
c}$ and $\xi_c$ starts with a linear in $X^a$ terms, the equation
\eq{kil} can be used to obtain restrictions on the coefficients
$P^a_b$, $P^a_{bc}$, and $P^a_{bcd}$ in \eq{S}.  
Simple calculations give
\be
P^1_1=P^2_2=0\hhh P^1_2=-P^2_1=q\hhh
P^a_{bc}=P^a_{bcd}=0\,,
\ee
\be\n{ab}
q k_1(k_1-k_2)=qk_2(k_1-k_2)=0\, .
\ee
If the Killing vector does not vanish identically then $q\ne0$ and the
equations \eq{ab} imply that $k_1=k_2$. This contradicts to the assumption
of the existence of the embedding with $K_0=k_1 k_2<0$. 

\section{Embedding of a 2D surface with $K_0<0$ in $\E^4$}

Increasing the number of dimensions of the flat space from 3 to 4
makes it possible to find an isometric embedding of 2D manifolds with
$K_0<0$.  Denote by $(X,Y,Z,R)$ Cartesian coordinates in $\E^4$ and
determine the embedding by equations
\be\n{cone}
X={r\over \Phi_0} \xi(\psi)\hhh
Y={r\over \Phi_0} \eta(\psi)\hhh
Z={r\over \Phi_0} \zeta(\psi)\, ,
\ee
\be\n{R}
R=R(r)\, ,
\ee
where $0\le \psi\le 2\pi$, and functions $\xi$, $\eta$ and $\zeta$ obey the
condition
\be
\xi^2(\psi)+\eta^2(\psi)+\zeta^2(\psi)=1\, .
\ee
In other words, ${\bf n}=(\xi,\eta,\zeta)$ as a function of $\psi$ is
a line on a unit sphere $S^2$. We require that this line is a smooth
closed loop (${\bf n}(0)={\bf n}(2\pi)$) without self-intersections. 
Since a loop on a unit sphere allows continuous deformations
preserving its length, there is an ambiguity in the choice of functions
$(\xi,\eta,\zeta)$.

We denote
$\Phi=({\xi_{,\psi}}^2+{\eta_{,\psi}}^2+{\zeta_{,\psi}}^2)^{1/2}$
then
\be
2\pi \Phi_0=\int_0^{2\pi} d\psi \Phi(\psi)\, .
\ee
is the length of the loop.
Instead of the coordinate $\psi$ it is convenient to use a new
angle coordinate $\phi$ which is proportional to the proper
length of a curve $r=$const 
\be
\phi =\Phi_0^{-1} \int_0^{\psi} d\psi' \Phi(\psi')\, .
\ee
The coordinate $\phi$ is a monotonic function of $\psi$ and for $\psi
=0$ and $\psi=2\pi$ it takes values $0$ and $2\pi$, respectively.

Equations \eq{cone} give  the embedding in $\E^3$ of a linear surface
formed by straight lines passing through $r=0$. This surface has
$K=0$ outside the point $r=0$ where, in a general case, it has a
cone-like singularity  with the angle deficit $2\pi (1-\Phi_0)$. 

\begin{figure}[tp]
\begin{center}
\includegraphics[height=3cm,width=4.5cm]{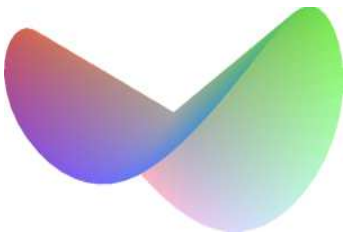}
\caption{The surface shown at this picture is formed by straight lines
passing through a point $r=0$. Its Gauss curvature vanishes. The
surface has a cone singularity at $r=0$ with a negative angle deficit.}
\label{f2}
\end{center}
\end{figure}

We shall use the embedding \eq{cone}--\eq{R} for
the case when the angle deficit is negative. In this case
one can use, for example, the following set of
functions
\be\n{hat}
\xi=\cos \psi/F\hhh
\eta=\sin \psi /F \hhh
\zeta=a \sin(2\psi)/F\, ,
\ee
\be\n{hat1}
F=\sqrt{1+a^2x^2}\hh x=\sin^2(2\psi)\, .
\ee
This embedding for the functions $(\xi,\eta,\zeta)$ defined by
\eq{hat}--\eq{hat1} is shown in Figure~\ref{f2}.

\begin{figure}[tpb]
\begin{center}
\includegraphics[height=3cm,width=5cm]{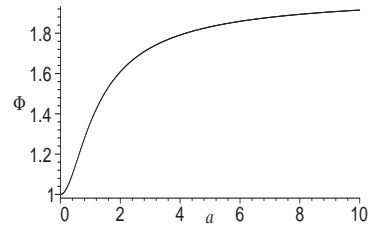}
\caption{$\Phi_0$ as a function of the parameter $a$.}
\label{f3}
\end{center}
\end{figure}

For this choice
\be
\Phi=(1+4a^2-3a^2 x)^{1/2}(a^2x+1)^{-1}\, ,
\ee
\be
\Phi_0={1\over \pi} \int_0^1 {dx (1+4a^2-3a^2 x)^{1/2}\over
\sqrt{x(1-x)} (a^2x+1)}\,  .
\ee
Calculations give
\be
\Phi_0={8\over \pi \sqrt{1+4a^2}}\left[
(1+a^2)\Pi(-a^2,k)
-{3/4}
K(k)\right]\, ,
\ee
\be
k=\sqrt{3}a/(1+4a^2)\, .
\ee
Here $K(k)$ and $\Pi(\nu,k)$ are complete elliptic integrals of the
first and third kind, respectively. The function $\Phi_0$
monotonically increases from $1$ (at $a=0$) to $2$ (at $a\to \infty$)
(see Fig.~\ref{f3}).

The induced metric for the embedded $2D$ surface defined by
\eq{cone}--\eq{R} is
\be\n{ind}
dl^2=[\Phi_0^{-2}+ (dR/dr)^2] dr^2+r^2 d\phi^2\, .
\ee
If the angle deficit is positive ($\Phi_0<1$), the pole point $r=0$
in the metric (\ref{cone}) remains a cone singular point for any
$R(r)$. For $\Phi>1$ (the negative angle deficit), the pole-point
$r=0$ in the metric \eq{ind} is regular if $(dR/dr)_0^2=1-\Phi_0^2$.
By comparing (\ref{VV}) and (\ref{ind}) one obtains
\be\n{emb}
r=f^{1/2}\hhh 
({dR/dr})^2=(V-\Phi_0^{-2})\, .
\ee
This relation gives the following equation relating $R(\mu)$ with $f(\mu)$
\be\n{Rp}
R'=(1- {f'}^2/(4\Phi_0^2))^{1/2} f^{-1/2}\, .
\ee
It is easy to check that $R''=0$ at points where $f'=0$. In order
$R'$ to be real, the following condition must be valid  $\Phi_0\ge
{1\over 2}\max_{\mu\in (-1,1)}|f'(\mu)|$. At a point where $|f'|$
reaches its maximum the quantity $f''=-2K$ vanishes. Thus it is
sufficient to require that $\Phi_0$ is greater or equal to the values
of $|f'|$ calculated at the points separating regions with the
positive and negative Gauss curvature.

\section{Embedding of the surface of the Kerr-Newman horizon in $\E^4$}

The surface geometry of the Kerr-Newman black hole is described by the
metric $ds^2=N^2 dl^2$, where
\be\n{ds}
dl^2=(1-\beta^2 \sin^2\theta)\, d\theta^2+\sin^2\theta [1-\beta^2
\sin^2\theta]^{-1}d\phi^2 \, ,
\ee
\be
N=(r_+^2+a^2)^{1/2}\hh \beta=a(r_+^2+a^2)^{-1/2}\, .
\ee
Here $0\le \theta \le \pi$ and $0\le \phi\le 2\pi$. The metric $dl^2$
is normalized so that the area of the surface with this metric is
$4\pi$.  In the coordinates $\mu=\cos\theta$ the metric \eq{ds}
takes the form \eq{m} with
\be
f(\mu)=(1-\mu^2)[1-\beta^2(1-\mu^2)]^{-1}\, .
\ee

\begin{figure}[tpb]
\begin{center}
\includegraphics[height=3cm,width=5cm]{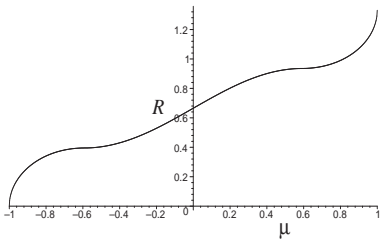}
\caption{Plot for $R$ as a function of $\mu$ for $\beta=0.7$.}
\label{f4}
\end{center}
\end{figure}

	For the black hole with mass $M$, charge
$Q$ and the angular momentum $J=Ma$
\be
r_+=M-(M^2-a^2-Q^2)^{1/2}\, . 
\ee
The rotation parameter $a$ and mass $M$ can be written in terms
of the distortion parameter $\beta$ as follows
\be
a=\beta N\hh 
M={1\over 2} N (1-\beta^2)^{-1/2} (1+Q^2/N^2)\, .
\ee 
The condition $M^2\ge a^2+Q^2$ for given parameters $N$ and $\beta$
requires that \cite{Smarr}
\be
0\le Q\le N (1-\beta^2)^{1/2}\,,
\ee
\be
{1\over 2} N (1-\beta^2)^{-1/2}\le M\le N  (1-\beta^2)^{1/2}\, .
\ee
The distortion parameter has its maximal value
$\beta_{max}=1/\sqrt{2}$ for $Q=0$.
The Gauss curvature of the surface with the metric \eq{ds} is
\be
K=[1-\beta^2(1+3\mu^2][1-\beta^2(1-\mu^2)]^{-3}\, .
\ee
For ${1\over 2}<\beta\le {1\over \sqrt{2}}$ the Gauss curvature is
negative in the vicinity of poles in the region $ \mu_c\le |\mu|\le
1$
\be
\mu_c= (1-\beta^2)^{1/2}(\sqrt{3}\beta)^{-1}\, .
\ee
At $|\mu|=\mu_c$ the Gauss curvature vanishes. As it was shown earlier,
at this point $|f'|$ has its  maximum 
\be
|f'|_{\max}=|f'|_{\mu_c}= {3\sqrt{3}\over 8 \beta (1-\beta^2)^{3/2}}\, ,
\ee
and one must choose the parameter $\Phi_0$ so that $\Phi_0\ge
{1\over 2}|f'|_{\max}$. Simplest possible choice is
\be\n{cond}
\Phi_0={1\over 2} |f'|_{\mu_c} \, .
\ee

\begin{figure}[t]
\begin{center}
\bigskip
\includegraphics[height=3.5cm,width=4.4cm]{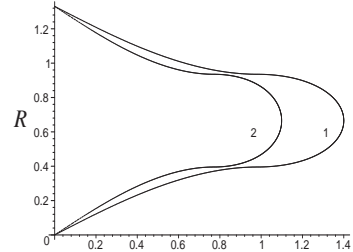}
\caption{Plot for $R$ as a function of $r$ (curve $1$) and $\rho$
(curve $2$) for $\beta=0.7$.}
\label{f5}
\end{center}
\end{figure}

Using \eq{cond} and integrating the equation \eq{Rp} one determines
$R$ as a function of $\mu$. A plot of this function for $\beta=0.7$
is shown in Figure~\ref{f4}. Plot~1 at Figure~\ref{f5} shows $R$ as a
function of $r$ for the same values of $\beta$.

\begin{figure}[tpb]
\begin{center}
\hfill
\includegraphics[height=2.7cm,width=2.7cm]{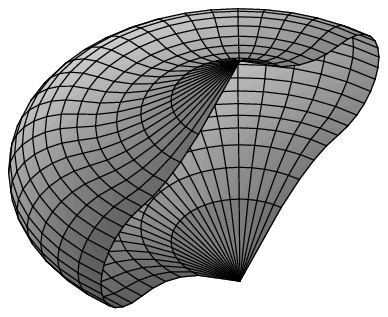}\hfill 
\includegraphics[height=2.7cm,width=2.7cm]{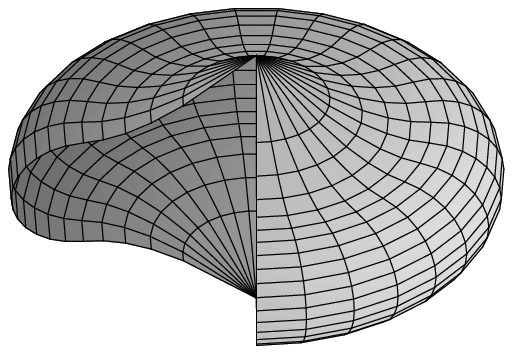}\hfill {  }
\caption{By gluing these two figures along their edges one obtains a
$2D$ surface without angle deficits  and isometric
to the  surface of a rotating black hole 
($\beta=0.7$.)}
\label{f6}
\end{center}
\end{figure}

The metric \eq{ind} can  also be written in the form
\be\n{rho}
dl^2=[1+(dR/d\rho)^2] d\rho^2+ \Phi_0^2\rho^2d\phi^2\, ,
\ee
where $\rho=r/\Phi_0$. Plot~2 at Figure~\ref{f5} shows $R$ as a
function of $\rho$ for
$\beta=0.7$. The metric \eq{rho} coincides  locally with the
metric on the revolution surface determined by the equation
$R=R(\rho)$ in $\E^3$. This does not give a global isometric embedding
since the period of the angle coordinate is $2\pi \Phi_0$. This
surface can be obtained by gluing two figures shown in
Figure~\ref{f6} along their edges. For the left figure $\phi$
changes from $0$ to $\pi$, while for the right one it changes from
$\pi$ to $2\pi \Phi_0$.

\section{Concluding remarks}

We demonstrated that a surface of a rapidly rotating black hole,
which cannot be isometrically embedded in $E^3$, allows such a global
embedding in $E^4$. To construct this embedding one considers first a
$2D$ surface in $E^3$ formed by straight lines passing through one
point (r=0) which has a cone singularity at $r=0$ with negative angle
deficit. Its Gauss curvature outside $r=0$ vanishes. Next element of
the construction is finding a function $R(r)$. The revolution surface
for this function in $E^3$ has a positive angle deficit at $r=0$. By
combining these two maps in such a way that positive and negative
angle deficits cancel one another, one obtains a regular global
embedding in $E^4$. This construction can easily be used to find the
embedding in $E^4$ of surfaces of rapidly rotating stationary black
holes distorted by an action of external forces or fields, provided
the axial symmetry of the spacetime is preserved. An interesting
example is a case of a rotating black hole in a homogeneous at
infinity magnetic field  directed along the axis of the rotation (see
e.g. \cite{wk,wkd,kd}.

\noindent 
\section*{Acknowledgments} 
\noindent 
This work was supported by the Natural Sciences and Engineering
Research Council of Canada and by the Killam Trust.



\begin{thebibliography}{}

\bibitem{jan} M. Janet, Ann. Soc. Polon. Math. {\bf 5}, 38 (1926).

\bibitem{car} E. Cartan,  Ann. Soc. Polon. Math. {\bf 6}, 1 (1927).

\bibitem{nash} J. Nash, Annals of Math. {\bf 63}, 20 (1956).

\bibitem{ber} M. Berger, {\em A Panoramic View of Riemannian
Geometry}, Springer-Verlag, Berlin-Heidelberg-New York, (2003).

\bibitem{pog} A. V. Pogorelov, Dokl. Akad. Nauk SSSR, {\bf 198}, 42
(1971); English translation in Soviet Math. Dokl. {\bf 12}, 729
(1971).

\bibitem{ny} N. Nadirashvili and Y. Yuan, {\em Counterexamples for
Local Isometric Embedding}, Eprint math.dg/0208127.

\bibitem{Smarr} L. Smarr, Phys. Rev.  {\bf D7}, 289 (1973).


\cqg{pnl}{N. Pelavas, N. Neary, and K. Lake}{18}{1319}{2001}

\cqg{bas}{M. Bondarescu, M. Alcubierre, and E. Seidel}{19}{375}{2002}


\bibitem{pol} This example demonstrates that the Proposition 4.1
of the paper \cite{str} is wrong. 

\bibitem{str} J. Stryla, Class. Quant. Grav. {\bf 15}, 2303 (1998).

\pr{wk}{W. J. Wild and R. M. Kerns}{21}{332}{1980}

\pr{wkd}{W. J. Wild, R. M. Kerns, and W. F. Drish, Jr.}{23}{829}{1981}

\pr{kd}{R. Kulkarni and N. Dadhich}{33}{2780}{1986}




\end{thebibliography}
\end{document}